# Quantum Melting of a Disordered Wigner Solid

Ziyu Xiang[1,2,3,†], Hongyuan Li[1,2,3,†], Jianghan Xiao[1,2,3,†], Mit H. Naik[1,3], Zhehao Ge[1], Zehao He[1], Sudi Chen[1,3,7], Jiahui Nie[1], Shiyu Li[1], Yifan Jiang[1], Renee Sailus[4], Rounak Banerjee[4], Takashi Taniguchi[5], Kenji Watanabe[6], Sefaattin Tongay[4], Steven G. Louie[1,3], Michael F. Crommie[1,3,7,*] and Feng Wang[1,3,7,*]

[1]Department of Physics, University of California at Berkeley, Berkeley, CA, USA.

[2]Graduate Group in Applied Science and Technology, University of California at Berkeley, Berkeley, CA, USA.

[3]Materials Sciences Division, Lawrence Berkeley National Laboratory, Berkeley, CA, USA.

[4]School for Engineering of Matter, Transport and Energy, Arizona State University, Tempe, AZ, USA.

[5]International Center for Materials Nanoarchitectonics, National Institute for Materials Science, Tsukuba, Japan

[6]Research Center for Functional Materials, National Institute for Materials Science, Tsukuba, Japan

[7]Kavli Energy Nano Sciences Institute at the University of California Berkeley and the Lawrence Berkeley National Laboratory, Berkeley, CA, USA.

† These authors contributed equally

* To whom correspondence should be addressed: crommie@berkeley.edu, fengwang76@berkeley.edu

**Abstract:** The behavior of two-dimensional electron gas (2DEG) in extreme coupling limits are reasonably well-understood, but our understanding of intermediate region remains limited. Strongly interacting electrons crystalize into a solid phase known as the Wigner crystal at very low densities[1–4], and these evolve to a Fermi liquid at high densities[5]. At intermediate densities, however, where the Wigner crystal melts into a strongly correlated electron fluid[3,6–11] that is poorly understood partly due to a lack of microscopic probes for delicate quantum phases. Here we report the first imaging of a disordered Wigner solid and its quantum densification and quantum melting behavior in a bilayer $MoSe_2$ using a non-invasive scanning tunneling microscopy (STM) technique. We observe a Wigner solid with nanocrystalline domains pinned by local disorder at low hole densities. With slightly increasing electrostatic gate voltages, the holes are added quantum mechanically during the densification of the disordered Wigner solid. As the hole density is increased above a threshold ($p \sim 5.7 \times 10^{12} \, cm^{-2}$), the Wigner solid is observed to melt locally and create a mixed phase where solid and liquid regions coexist. With increasing density, the liquid regions gradually expand and form an apparent percolation network. Local solid domains appear to be pinned and stabilized by local disorder over a range of densities. Our observations are consistent with a microemulsion picture of Wigner solid quantum melting where solid and liquid domains emerge spontaneously and solid domains are pinned by local disorder[6,10,11].

Understanding strongly correlated electron phenomena arising from the interplay between the Coulomb energy and the electron kinetic energy is a central topic in condensed matter physics [2,12–16]. Two-dimensional electron gas (2DEG) of very high densities are well-described by a weakly interacting Fermi liquid theory [5] whereas in the low-density limit electrons solidify into a Wigner crystal where Coulomb energy dominates over kinetic energy[1,17,18]. At intermediate densities, however, the competition between Coulomb energy and kinetic energy is more pronounced and drives the emergence of exotic strongly correlated phases[19,20]. Great experimental and theoretical effort has been made to understand the quantum melting of Wigner crystals and the resulting correlated electron phenomena at intermediate electron densities [3,6–11,19,21,22]. Striking non-Fermi liquid transport behavior, including an insulator-to-metal transition and anomalous temperature- and magnetic field-dependent resistivity, have been observed experimentally in various 2DEG systems [19,21–23]. While different theoretical pictures have been proposed to describe the experimental data [3,6–11], no consensus on the proper description of Wigner crystal quantum melting and resulting correlated electron states at intermediate electron densities has been reached. For significant further understanding of the microscopic nature of Wigner solid quantum melting, new experimental measurements capable of probing correlated electronic structure at both atomic and mesoscopic scales are necessary.

Here we describe the use of a non-invasive imaging technique to visualize a Wigner solid of holes, including its densification and quantum melting, in bilayer $MoSe_2$ (bi$MoSe_2$). At low hole density, we observe a disordered Wigner solid that exhibits a local triangular lattice but no long-range crystalline order due to point defects that pin the Wigner solid and induce nanocrystalline domains. The Wigner solid exhibits an unusual densification behavior as the hole density is increased: instead of a continuous contraction of the lattice constant as expected for a perfect Wigner crystal, the holes appear to evolve into a quantum superposition state. Above a critical hole density of $p = 5.7 \times 10^{12}\ cm^{-2}$, the Wigner solid starts to melt and enters a mixed phase with coexisting solid and liquid regions. The solid regions are typically pinned by defect sites and persist up to a local hole density of $7.0 \times 10^{12}\ cm^{-2}$. Defect-free regions, in contrast, convert to the liquid phase even at the same hole densities as the solid regions. The solid-liquid mixed phase is not due to a spatially varying hole density from long-range potential fluctuations. Instead, this behavior can be explained by a microemulsion picture where regions of solid and liquid phase emerge spontaneously during the quantum melting process [19,23] and solid regions are stabilized by local disorder[24]. For such a liquid-solid microemulsion the liquid regions expand gradually to form a percolation network as hole density is increased. Such behavior might explain insulator-to-metal transitions observed in electrical transport studies[10,11,23].

The ratio between Coulomb energy and kinetic energy in a 2DEG is characterized by a single dimensionless parameter $r_s = \frac{a_0}{a_B} = \frac{m^* e^2}{4\pi\varepsilon\hbar^2 \sqrt{\pi n}}$, , where $a_0$ is the average interparticle distance, $a_B$ is the effective Bohr radius, n is the electron/hole density, $\varepsilon$ is the effective dielectric constant, and $m^*$ is the electron effective mass. Quantum Monte Carlo simulations suggest that 2D Wigner crystals melt into a liquid phase around $r_s^* \approx 38$ [3,20,25]. 2D transition metal dichalcogenide semiconductors provide a rich platform to explore Wigner crystal physics[16,18,26]. We chose the 2D hole gas (2DHG) in biMoSe$_2$ as a model system to explore Wigner solid phenomena due to the large hole effective mass $m^* \approx -1.26 m_e$ predicted by our *ab initio* calculation (see supplementary section 3) which helps to increase $r_s$. Such an increased $r_s$ facilitates the formation of Wigner solids over a large hole density range and helps them to be more robust against temperature fluctuation and weak disorder. Fig. 1a illustrates the experimental setup incorporating biMoSe₂ into a van der Waals heterostructure device stack. The hole density in the biMoSe₂ is tuned by applying a bottom gate voltage $V_{BG}$ between a graphite bottom gate (BG) and the biMoSe₂. A graphene nanoribbon (GNR) array is specially prepared as the electrical contact to biMoSe₂, thus reducing contact resistance and facilitating effective gating. Fig. 1b displays an optical micrograph of the biMoSe₂ heterostructure device with biMoSe₂, GNR, and BG delineated by green, blue, and red dashed lines, respectively.

We employ a non-invasive valence band edge (VBE) tunnel current measurement technique[27] that enables Wigner solids to be probed in hole-doped biMoSe₂ with minimal tip perturbation. The schematic in Fig. 1c outlines the basic principle of the measurement by showing the band alignment between the STM tip and hole-doped biMoSe₂. Due to the non-negligible work function difference between the STM tip (made of Pt/Ir) and biMoSe₂, the tip perturbation cannot be neglected in most cases and often has a fatal impact on fragile correlated electrons. To overcome this issue, we tune the sample-tip bias voltage, $V_{bias}$, so that the vacuum energy levels of both the tip and biMoSe₂ are aligned. This minimizes the local electric field near the tip apex and ensures that the integrity of the Wigner solid is maintained (see supplementary section 1 for details). The alignment of the TMD and tip vacuum levels can be achieved by moving the tip chemical potential within the biMoSe₂ bandgap. In this case the tunnel current arises exclusively from doped holes at the biMoSe₂ VBE without impact from higher-energy states. The VBE tunnel current thus directly reflects the spatial distribution of doped holes similar to previous demonstrations of VBE current mapping of Wigner molecules[27].

Figs. 1d shows the tunnel current (I-V) characteristic on a log scale as a function of $V_{BG}$ for hole-doped biMoSe₂ while maintaining a relatively large tip-sample separation (determined by the setpoint condition $V_{bias}$ = 3V, $I_{sp}$ = 20pA, and $h_{tip}$ = -30pm, T = 5.4K, see methods).

Negligible tunnel current occurs for -0.5V < $V_{bias}$ < 2V which corresponds to the biMoSe₂ semiconducting band gap (the valence (VBE) and conduction band (CBE) edges are marked with dashed lines). The band gap shows an apparent increase as $V_{BG}$ is increased. This is likely due to an artifact arising from increasing contact resistance between the biMoSe₂ and GNR contact array at lower hole doping which leads to a part of bias voltage dropping at the contact rather than at the tip-sample gap. At large tip-sample separation the VBE tunnel current is lower than the measurement noise floor in the bandgap region and is not observable. When the tip-sample distance is reduced (setpoint condition: $V_{bias}$ = 3V, $I_{sp}$ = 20pA, and $h_{tip}$ = -90pm), however, VBE tunnel current starts to appear in the band gap, as denoted by the white arrow in Fig. 1e.

Mapping the VBE tunnel current allows us to spatially resolve the Wigner solid within bilayer MoSe₂. Fig. 2a shows the VBE tunnel current map obtained at a hole density of $p = 5.3 \times 10^{12} cm^{-2}$ (with V$_{BG}$ = -8V, $V_{bias}$ = 1.9V to minimize tip perturbation, see supplementary for methods on determining $V_{bias}$ with minimized tip perturbation). Each bright dot here represents a localized hole in biMoSe₂. Figure 2b shows the corresponding topographic image. The bright defects in Fig. 2b correspond to charged defects, which act as local pinning centers in the Wigner solid of Fig. 2a (Figure 2b also contains many weaker features arising from isovalent defects. See supplementary information for details)[28–30]. The Wigner solid exhibits a nanocrystalline phase where different nanocrystalline domains are present but no long-range crystalline order exists due to defect-induced disorder. Figs. 2d-2f display the fast Fourier transform (FFT) patterns corresponding to regions (i), (ii), and (iii) in Fig. 2a. Each shows a local hexagonal lattice with six distinct diffraction spots (denoted by red solid dots), but the lattice orientation is random in different regions. The orientational coherence of the Wigner solid is clearly disrupted by disorder in the biMoSe₂ semiconductor.

We employ the Delaunay triangulation and Voronoi cell partition method to systematically analyze the disordered Wigner solid. The Voronoi cell methodology, commonly used in defining the vortex lattice in layered superconductors [31,32], allows for quantitative analysis of the lattice structure of disordered crystals structure. Figure 2c shows the Voronoi cell result of our disordered Wigner solid (see supplementary section 2 for details). Each vertex corresponds to a single hole, and each edge connects a hole to a nearest neighbor. In an ideal Wigner crystal, every hole would have six nearest neighbors, but in the actual disordered Wigner solid a significant fraction of holes has five or seven nearest neighbors due to the large "strain" induced by local disorders. Consequently, each hole can be identified as the nucleus of either a pentagonal, hexagonal, or heptagonal cell. Of the 237 holes shown in Fig. 2c the number of hexagonal, pentagonal, and heptagonal cells are 151, 44, and 41, respectively. We have color coded the Voronoi cells so that blue dashed lines connect the vertices of pentagons and heptagons, while solid lines connect to hexagons. The three different colors (light blue, light orange, and dark red) of solid lines are applied to trace the lattice orientation of the crystalline domains. Pentagons and heptagons are observed to frequently appear in pairs, or defect lines. Such defects separate the hexagonal regions (solid lines) into randomly oriented nanocrystalline domains.

We study the evolution of the disordered Wigner solid as a function of the hole density by varying the gate voltage $V_{BG}$. Figs 2a, 2g-i show the VBE tunnel current maps for biMoSe$_2$ as the hole density is increased from $4.2 * 10^{12}$ cm$^{-2}$ to $7.0 * 10^{12}$ cm$^{-2}$ (see supplementary section 1 for discussion of $V_{bias}$ dependence). For low charge density ($V_{BG} \geq -8V$) the holes remain well-separated and form a solid phase. As $V_{BG}$ is reduced in this range the Wigner solid becomes denser and the spatial extent of the hole wavefunction shrinks. As $V_{BG}$ is reduced beyond the threshold of $V_{BG}$ = -8V, regions of the Wigner solid become blurred and no longer exhibit well-separated holes. This behavior indicates the onset of local melting of the Wigner solid. As $V_{BG}$ is further reduced the increased hole density causes the melted regions to expand and eventually form a percolation network at very high hole density ($p \sim 7.0 * 10^{12}$ cm$^{-2}$).

The local mechanism by which the density of hole increases as $V_{BG}$ is decreased for our disordered Wigner solid is very different from the smooth lattice constant variation expected for an ideal Wigner crystal. We observe instead a highly local process that we term "quantum densification". Here individual holes are added quantum mechanically into local regions constrained by local disorders, where partially delocalized hole wavefunction can exist during transition between two well-defined solid lattice configurations.

The quantum densification process can be observed in Fig. 3 which shows a small region of the Wigner solid imaged under conditions when the vacuum levels near the tip and sample surface are nearly in alignment. The dashed white circle in Fig. 3a outlines the region where the most dramatic changes occur during the quantum densification process as the local hole number increases from four to five as $V_{bias}$ is increased in Figs. 3a-3c. The red dots in every panel of Fig. 3 mark the same locations of the holes in Fig. 3a so that changes in hole positions can be easily seen. The white arrow in Fig. 3b points to the region where the new charge carrier comes in as the $V_{bias}$ is raised from $V_{bias}$ = 1.9V to $V_{bias}$ = 2.0V. A new coalescence of charge is seen to gather in the region indicated by the white arrow while the charge that was previously at the location of the red dot just above the arrow is seen to slightly rise. All other holes remain close to the locations of the red dots. The process continues in Fig. 3c where the sample bias is raised to $V_{bias}$ = 2.1V. The new charge carrier is now fully incorporated in the Wigner solid. This process continues in Fig. 3d where now $V_{BG}$ has been reduced to $V_{BG}$ = -8V, causing a new coalescence of hole gathering around the location indicated by the white arrow. When $V_{bias} = 1.9V$, six holes now exist in the center region (in Fig. 3e) where previously (in Fig. 3a) only four holes resided. The process continues in Figs. 3e, f where an increase in $V_{bias}$ causes a new coalescence of charge to appear at the region indicated by the arrow in Fig. 3e and the new charge further incorporated in the Wigner solid in Fig. 3f. The positions of the other holes only shift slightly.

During the transition between the integer-hole-number states under quantum densification, there is notable elongation or delocalization of the hole wavefunctions. Similar behavior is observed in other local regions during hole densification of the Wigner solid (see supplementary section 6 for additional data). This behavior is a manifestation of the quantum mechanical nature of the hole lattice, where quantum tunneling and delocalization of isolated holes can be significant under suitable conditions even in the solid phase. In contrast, atom positions are almost always localized in disordered natural solids due to the much heavier mass of atoms.

Next we describe the quantum melting behavior of the Wigner solid. Figure 4 shows this melting process as the hole density $p$ ranges from $5.3 * 10^{12}$ cm$^{-2}$ to $7.0 * 10^{12}$ cm$^{-2}$. At the highest hole density, we clearly observe a mixed phase composed of solid regions with a local hole lattice and liquid regions with a smooth hole density distribution. We define liquid (solid) regions as areas where the local tunnel current variation is lower (higher) than 28 percent over one Voronoi-cell size (see supplementary section 7 for more details). The boundaries between solid and liquid regions are delineated by two-toned solid lines in Fig. 4. The low-contrast liquid regions where holes are delocalized and have more widespread wavefunctions (close to the white boundary line), can be clearly distinguished from the high-contrast solid regions where hole positions remain more highly localized (close to the red boundary line). The liquid-solid mixed phase extends over a rather broad hole density range between $5.7 \times 10^{12}\ cm^{-2}$ to $7.0 \times 10^{12}\ cm^{-2}$. They correspond to the $r_s$ value of 21.8 to 19.7 if we use a hole effective mass of $-1.26 m_e$ in bilayer MoSe2 based on the *ab initio* calculation (see supplementary information). This range is smaller than the theoretically predicted Wigner crystal melting density at $r_s \approx 38$[3,20,25]. However, there is no experimentally measured hole effective mass. Previous studies that measured effective mass in 2D transition metal dichalcogenide semiconductors can be much larger than theoretical predictions, so there is a large uncertainty in the estimated $r_s$[33].

At $5.7 \times 10^{12}\ cm^{-2}$ only small bubbles of the liquid phase are present (Fig. 4b). As the hole density is steadily increased the liquid bubbles gradually expand (Fig. 4c) and coalesce with neighboring liquid regions, finally forming an interconnected percolation network (Fig. 4d). The percolation behavior observed here for the liquid region could potentially explain insulator-to-metal transitions seen for Wigner solids in previous transport studies[10,11,24].

The mixed phase quantum melting described here for disordered Wigner solids can be theoretically explained through two very different physical mechanisms. In one mechanism, the disordered 2D material has long-range potential fluctuations that lead to spatially varying hole density regions within the sample. When the average hole concentration is close to the critical density of the Wigner solid then low-density regions will be in the solid phase ($r_s > r_s^*$) while the high-density regions will be in the liquid phase ($r_s < r_s^*$). Here solid and liquid regions are directly correlated with the long-range potential disorder. In the second mechanism a microemulsion of liquid and solid regions emerges spontaneously due to the thermodynamic melting of the Wigner solid [6,10]. In an ideal 2D material free of disorder, the liquid and solid regions would be distributed randomly in a dynamic fashion. In a real material with disorder, however, solid regions can be pinned by disorder. This gives rise to a stable mixed phase where solid regions are correlated with disorder but not with hole density variations.

To distinguish these two mechanisms, we characterize the effect of the disorder induced potential landscape on the hole density distribution. If long-range potential fluctuations are the dominant effect, then they should lead to persistent hole density fluctuation patterns at all gate voltages. To test this idea, we define the liquid and solid regions in the mixed phase at $p = 7 \times 10^{12} cm^{-2}$ (i.e. $V_{BG}$ = -12V, $V_{bias}$ = 1.5V) as region I and region II, respectively, and compare the average hole density in these two regions at different gate voltages. We then use the inverse of the Voronoi cell area ($1/A_{Voronoi}$) to define the local hole density for Wigner solid states at $p < 5.5 \times 10^{12} cm^{-2}$ (see section 8 of the supplementary text for details). Figure 4e shows the hole density distribution based on the Voronoi cell analysis at $V_{BG}$ = -8V, $V_{bias}$ = 1.9V, which is overlayed with the boundaries of region I and region II. Significant spatial variations are seen in the hole density at the unit cell level but there are no pronounced long-range variations. The calculated average hole densities in region I and region II are identical within an experimental uncertainty of 2% (see supplementary for details), so there is no correlation between the solid (liquid) regions in the mixed phase and decreased (increased) local hole densities. Instead, solid regions strongly correlate with local disorder, suggesting local pinning and stabilization of the solid phase by defects. Our data is consistent with the microemulsion model of Wigner solid quantum melting where solid and liquid regions spontaneously emerge and solid regions are anchored by local disorder.

In conclusion, we have successfully visualized the quantum melting of a Wigner solid where solid and liquid phases coexist. This is consistent with the microemulsion picture of how a disordered Wigner solid transitions to a strongly correlated Fermi liquid and may explain percolative metal-insulator transitions observed previously via transport measurement.

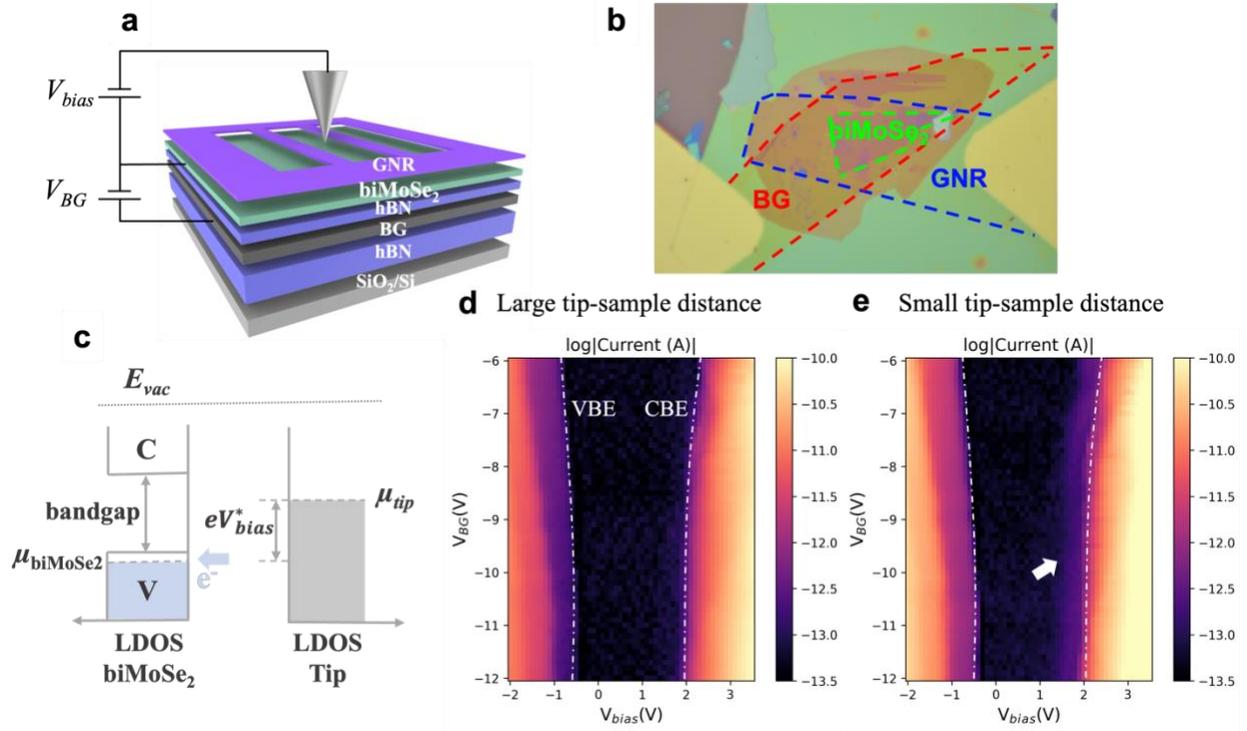

**Fig. 1 | Non-invasive valence band edge (VBE) tunnel current measurement of a bilayer MoSe₂ (biMoSe₂). a**, A schematic of the STM measurement setup for a gate-tunable biMoSe₂. The biMoSe₂ is placed on top of a 50nm thick hBN layer and a graphite substrate that defines the back gate (BG). A back gate voltage $V_{BG}$ is applied to control the charge carrier density in the biMoSe₂. A bias voltage $V_{bias}$ is applied to the biMoSe₂ relative to the STM tip to induce tunnel current. A graphene nanoribbon (GNR) array is placed on top of biMoSe₂ as the contact electrode. **b**, Optical microscope image of the device heterostructure. The biMoSe₂, GNR, and BG are outlined in green, blue, and red, respectively. **c**, The schematic energy diagram for the valence band edge (VBE) tunnel current measurement of hole-doped biMoSe₂. Here the chemical potential, $\mu_{biMoSe2}$, is near the valence band edge. When $\mu_{tip}$ is aligned within the band gap of biMoSe₂ then the tunnel current arises from doped holes at the VBE. $V_{bias}$ is tuned to roughly align the vacuum energy levels of the tip and biMoSe₂ so that the electric field near the tip apex is minimized. The gap between $\mu_{tip}$ and $\mu_{biMoSe2}$ is denoted $V_{bias}^*$ instead of $V_{bias}$ since the real bias on the tunneling junction $V_{bias}^*$ is reduced by the large contact resistance (i.e., the Schottky barrier). **d**, The tunnel current I-V characteristics as a function of $V_{BG}$ for hole-doped biMoSe₂ with a large tip-sample distance. The current is plotted on an absolute log scale. The valence band and conduction band edges are labeled with white dashed curves. **e**, The tunnel current I-V characteristics as a function of $V_{BG}$ for hole-doped biMoSe₂ with a small tip-sample distance, similar to **d**. The white arrow marks the VBE tunnel current. All STM data was acquired at T = 5.4K.

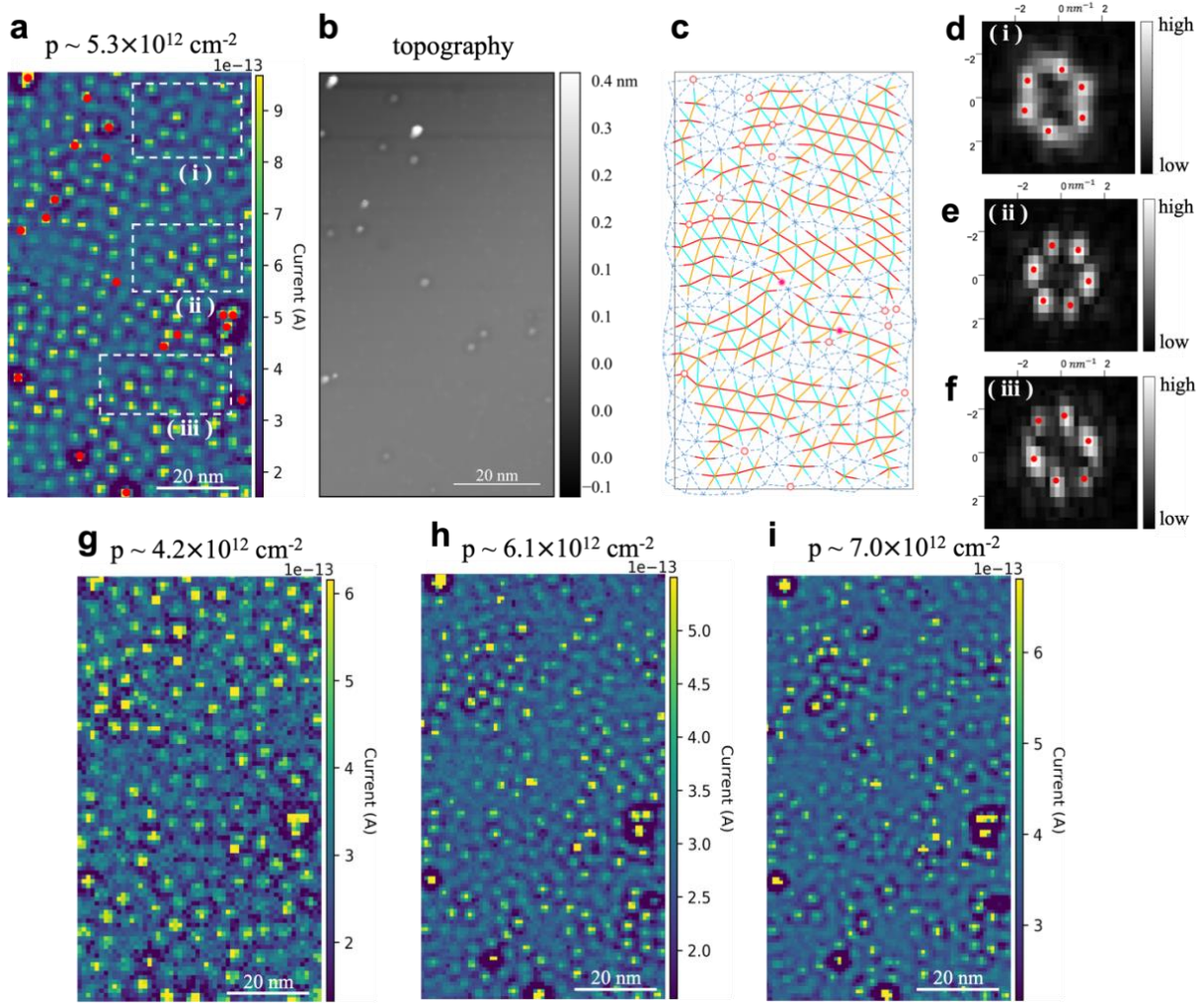

**Fig. 2 | Hole density dependence of a Wigner solid. a**, The VBE tunnel current map of a Wigner solid in a 60nm*100nm region ($V_{BG}$ = -8V, $V_{bias}$ = 1.9V). The bright dots in the VBE tunnel current maps correspond to individual holes in the Wigner solid. Charged defects are labeled in **a** with red solid dots. **b**, STM topographic image of the biMoSe$_2$ surface taken in the same region as **a, g-i**. Most charged defects in the VBE maps correspond to defects in the topography image. **d-f**, Fast Fourier transform (FFT) images for regions (i), (ii), and (iii) showed in **a** (a Nuttall window was used). The red solid dots mark the peaks in the FFT images. **c**, The crystal structure analysis of **a**. Each intersection marks a hole position calculated by applying Voronoi cell partition method (see supplementary for details). The points connected with blue dashed lines are holes that have 5 or 7 nearest neighbors, while the points connected with colored solid lines are holes that have 6 nearest neighbors. The red open circles mark charged defect positions shown in **a**. Solid red points mark defects without a trapped hole. **g-i,** The VBE tunnel current maps of a Wigner solid for different hole densities controlled using the bottom gate voltage $V_{BG}$. **g**, $V_{BG}$ = -6V, $V_{bias}$ = 2.2V. **h**, $V_{BG}$ = -10V, $V_{bias}$ = 1.8V. **i**, $V_{BG}$ = -12V, $V_{bias}$ = 1.5V.

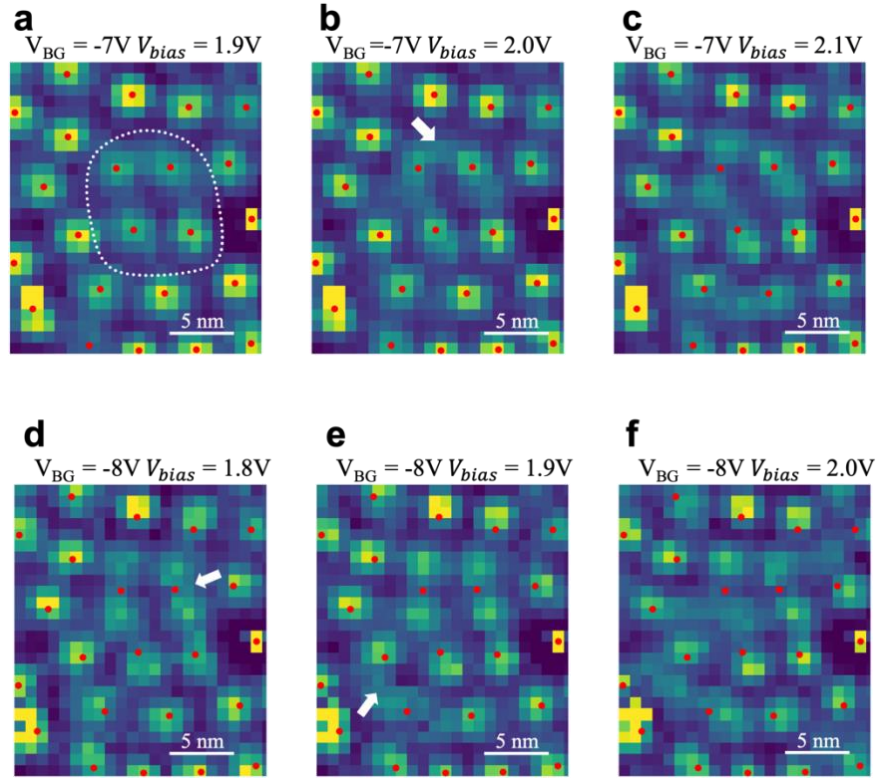

**Fig. 3 | Quantum densification of holes in a Wigner solid.** VBE tunnel current maps show a Wigner solid with hole density increasing in fine steps using both $V_{BG}$ and $V_{bias}$ (with negligible tip perturbation, see supplementary for details). **a**, $V_{BG}$ = -7V, $V_{bias}$ = 1.9V. **b**, $V_{BG}$ = -7V, $V_{bias}$ = 2.0V. **c**, $V_{BG}$ = -7V, $V_{bias}$ = 2.1V. **d**, $V_{BG}$ = -8V, $V_{bias}$ = 1.8V. **e**, $V_{BG}$ = -8V, $V_{bias}$ = 1.9V. **f**, $V_{BG}$ = -8V, $V_{bias}$ = 2.0V. Solid red dots mark the initial location of holes in **a** for all panels. Region marked by white dashed line in **a** show a local 4-hole configuration (a), which evolves into a well resolved 5-hole configuration (c) and 6-hole configuration (e) at higher densities. During the intermediate densities in (b) and (d), the wavefunction of local holes become partially delocalized (white arrows), a manifestation of the quantum nature of the Wigner hole solid.

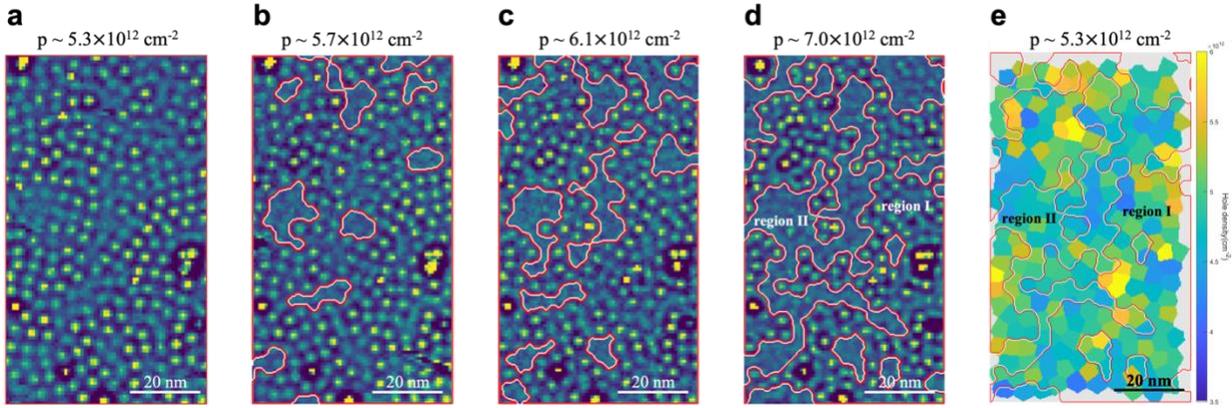

**Fig. 4 | Quantum melting of a Wigner solid.** The VBE tunnel current maps of a Wigner solid show a gradual melting at high charge densities. **a**, $V_{BG}$ = -8V, $V_{bias}$ = 1.9V. **b**, $V_{BG}$ = -9V, $V_{bias}$ = 1.8V. **c**, $V_{BG}$ = -10V, $V_{bias}$ = 1.8V. **d**, $V_{BG}$ = -12V, $V_{bias}$ = 1.5V. The boundary between solid and liquid phase is distinguished by its local contrast difference and is outlined by two-colored solid lines. The red side indicates solid phase while the white side indicates liquid phase. Wigner solid domains pinned by defects shrink with increasing charge density while liquid regions expand and increase percolative connectivity. We label the liquid phase and solid phase regions in **d** as region I and region II, respectively. **e**, The hole density distribution at $V_{BG}$ = -8V, $V_{bias}$ = 1.9V (yellow is high, blue is low). The overlaid solid lines delineate region I and region II as defined in d. The average charge densities in the two regions are identical within an experimental uncertainty of 2%.

**Corresponding Author**

* Email: crommie@berkeley.edu (M.C.) and fengwang76@berkeley.edu (F.W.).

**Author Contributions**

M.F.C., and F.W. conceived the project. Z.X., J.X., H.L., J.N., S.L., and Y.J. fabricated the heterostructure device. Z.X. and S.C. fabricated the shadow mask. Z.X., J.X. and H.L. performed the STM/STS measurement. M.H.N., and S.G.L. performed the DFT calculations of the $biMoSe_2$ effective mass. Z.X., J.X., H.L., M.F.C. and F.W. discussed the experimental design and analyzed the experimental data. R.S., R.B. and S.T. grew the $MoSe_2$ crystals. K.W. and T.T. grew the hBN single crystal. All authors discussed the results and wrote the manuscript.

**Notes**

The authors declare no financial competing interests.

**ACKNOWLEDGMENT**

The authors would like to acknowledge helpful discussion with Professor Steven Kivelson and Philip Kim. This work was primarily funded by the U.S. Department of Energy, Office of Science, Basic Energy Sciences, Materials Sciences and Engineering Division under Contract No. DE-AC02-05-CH11231 within the van der Waals heterostructure program KCWF16 (device fabrication, STM spectroscopy, theoretical analysis, and computations). Support was also provided by the National Science Foundation Award DMR-2221750 (surface preparation). This research used the Lawrencium computational cluster provided by the Lawrence Berkeley National Laboratory (Supported by the U.S. Department of Energy, Office of Basic Energy Sciences under Contract No. DE-AC02-05-CH11231). This research also used resources of National Energy Research Scientific Computing Center (NERSC), a U.S. Department of Energy Office of Science User Facility located at Lawrence Berkeley National Laboratory, operated under Contract No. DE-AC02-05CH11231. S. T. acknowledges primary support from U.S. Department of Energy-SC0020653 (materials synthesis), NSF CMMI1825594 (NMR and TEM studies on crystals), NSF DMR-1955889 (magnetic measurements on crystals), NSF ECCS2052527 (for bulk electrical tests), DMR 2111812, and CMMI 2129412 (for optical tests on bulk crystals). K.W. and T.T. acknowledge support from the JSPS KAKENHI (Grant Numbers 21H05233 and 23H02052) and World Premier International Research Center Initiative (WPI), MEXT, Japan.

## Methods

**Sample fabrication.** The biMoSe₂ device was fabricated using a micromechanical stacking technique[34]. A poly(propylene) carbonate (PPC) film stamp was used to stack all exfoliated 2D material layers. The 2D layers of the main heterostructure were picked up in the following order: hexagonal boron nitride (hBN) substrate, graphite, bottom hBN, bilayer MoSe₂, and lastly the graphene nanoribbon array which serves as a contact electrode for the biMoSe₂. The PPC film and stacked sample were peeled together, flipped over, and then transferred onto a Si/SiO₂ substrate (SiO₂ thickness 285nm). The PPC layer was subsequently removed using ultrahigh vacuum annealing at 340 °C, resulting in an atomically clean heterostructure suitable for STM measurements. A layer of 60nm thick Au and a 5nm thick Cr metal electrode was evaporated onto the sample using a shadow mask.

**Scanning tunneling microscopy (STM) valence band edge (VBE) tunnel current spectroscopy and mapping measurements.** Tunnel current I-V characteristics and VBE tunnel current mapping measurements[27] were performed under open-loop conditions with the tip height set by the following procedure: (1) Initial conditions were set by applying a bias voltage $V_{bias}$ = -3.0V with tunnel current I = 20 pA. This was achieved with the feedback loop engaged to ensure stable initial tip-sample separation. (2) The feedback loop was then deactivated and the tip lowered by 90pm to improve the signal-to-noise ratio for measurement. All measurements were performed at T=5.4K.

## Supplementary Materials

**Data availability**

The data supporting the findings of this study are included in the main text and in the Supplementary Information files, and are also available from the corresponding authors upon request.